\documentclass[aps,amsfonts,nofootinbib,superscriptaddress,preprint]{revtex4-1}
\usepackage[normalem]{ulem} 
\makeatletter
\makeatletter
\usepackage{tikz}
\usetikzlibrary{decorations.pathmorphing, shapes.misc, shapes}
\usepackage{amsmath}
\usepackage{mdframed} 
\usepackage{tikz-feynman}
\tikzfeynmanset{compat=1.1.0}
\usepackage{tikz-feynman} 
\tikzfeynmanset{compat=1.1.0} 
\usepackage{feynmp-auto}
\usepackage{cleveref}
\usepackage{amssymb}
\usepackage{xcolor} 
\usepackage{tcolorbox} 
\usepackage{amsbsy}
\usepackage{bm}
\usepackage{epsfig} 
\usepackage{color}
\usepackage{slashed}
\usepackage{soul}
\usepackage{comment}
\usepackage{auncial}
\makeatletter
\usepackage{appendix}
\usepackage{epsfig} 
\usepackage{ulem}
\usepackage{pgfplots}
\newcommand{\stkout}[1]{\ifmmode\text{\sout{\ensuremath{#1}}}\else\sout{#1}\fi}
\newcommand{\ee}{\end{equation}}
\newcommand{\bb}{\begin{equation}}
\newcommand{\eqb}{\begin{eqnarray}}

\newcommand{\eqf}{\end{eqnarray}}

\usepackage[T1]{fontenc}
\usepackage{comment}
\usepackage[normalem]{ulem} 
\makeatletter
\usepackage{xcolor}
\usepackage{mdframed}
\usepackage{amsmath}
\usepackage{amssymb}
\usepackage{appendix}
\usepackage{amsbsy}
\usepackage{tikz}
\usepackage{booktabs}
\usetikzlibrary{decorations.pathmorphing, shapes.misc, shapes}
\usepackage{amsmath}
\usepackage{tikz}
\usepackage{feynmp-auto}
\usepackage{cleveref}
\usepackage{bm}
\usepackage{epsfig} 
\usepackage{color}
\usepackage{slashed}
\usepackage{soul}
\usepackage{comment}
\usepackage[normalem]{ulem} 
\makeatletter
\usepackage{amsmath}
\usepackage{amssymb}
\usepackage{appendix}
\usepackage{amsbsy}
\usepackage{bm}
\usepackage{epsfig} 
\usepackage{color}
\usepackage{slashed}
\usepackage{soul}
\begin{document}
\title{Berry Phase in Non-Perturbative QED }
\author{J. Gamboa}
\email{jorge.gamboa@usach.cl}
\affiliation{Departamento de F\'{\i}sica, Universidad de Santiago de Chile, Santiago 9170020, Chile}
\begin{abstract}
We study QED$_4$ in the adiabatic approximation, incorporating global topological effects associated with the $U(1)$ Berry connection. The Berry phase accumulated by the fermionic vacuum is given by $\Delta \alpha = \oint_{\mathcal{C}} \gamma_5\, \mathcal{A}^{(n)}$, where $\mathcal{A}^{(n)}$ is a closed but non-exact one-form defined over the space of gauge configurations. This chiral holonomy induces an emergent vacuum angle that contributes non-perturbatively to the effective action. The partition function decomposes into topological sectors weighted by this geometric phase, analogous to quantum systems on multiply connected spaces. Our results reveal that even in Abelian gauge theory, the infrared regime can exhibit global effects beyond the reach of local or perturbative descriptions.
\end{abstract}

\maketitle

\vskip 0.25cm
\noindent {\bf Introduction}: Quantum electrodynamics (QED) is the first successful quantum field theory and remains one of the most precisely tested frameworks in physics. Among its celebrated triumphs is the theoretical prediction of the electron's anomalous magnetic moment, which agrees with the experiment to an extraordinary degree. Based on the Feynman diagram expansion in the coupling constant $e$, the perturbative sector of QED has produced a wealth of results with unmatched precision.

In contrast, the non-perturbative structure of QED is far less understood. While some important non-perturbative effects are well established—such as a) the chiral anomaly  \cite{jackiw,adler,fujikawa1,fujikawa2}, b) the Schwinger effect of vacuum pair production in strong electric fields \cite{Schwinger:1951nm}, and c) the non-perturbative formulation via lattice regularization \cite{wilson,suss,rev}—few additional phenomena have been identified in this regime.

Furthermore, the infrared sector of QED has long presented subtle challenges. The foundational work of Bloch and Nordsieck \cite{Bloch:1937pw} and Yennie et al \cite{frautschi}, later refined by Kinoshita \cite{Kinoshita:1962ur} and Lee and Nauenberg \cite{Lee:1964is}, established the necessity of summing over degenerate states to cancel infrared divergences, emphasizing the nontrivial structure of asymptotic states in QED.

Motivated by these foundational insights, we revisit the infrared sector of QED from a new perspective, inspired by the geometric phase concept introduced by M. Berry \cite{Berry:1984jv}. In this Letter, we explore whether a Berry phase—arising from the adiabatic evolution of fermionic states—can contribute non-perturbatively to the effective action of QED. This approach suggests a novel geometric structure associated with the chiral sector, with possible implications for the infrared behavior of QED beyond the perturbative expansion \footnote{In recent years, the infrared structure of QED has been studied from various perspectives, including the framework of asymptotic symmetries~\cite{stro}, as well as through other approaches such as those discussed in~\cite{otros1}.
}.
\vskip 0.25cm
\noindent {\bf {Developing the Idea}}: To develop the ideas outlined above, let us consider the generating functional of QED:
\begin{equation}
\mathcal{Z} = \int [\mathcal{D}A]\, \mathcal{D}\bar{\psi}\, \mathcal{D}\psi \, e^{-S[\bar{\psi}, \psi, A]},
\end{equation}
where \( [\mathcal{D}A] \) denotes the Faddeev–Popov gauge-fixed measure, and the fermion mass has been omitted as it is not relevant for the present discussion and 
\bb
S = \int d^4x \left( \frac{1}{4} F_{\mu\nu} F^{\mu\nu} + \bar{\psi}( \slashed D [A]) \psi \right), \label{sch2}
\ee
and $D_\mu [A]= \partial_\mu + ie A_\mu$.
\vskip 0.25cm
\noindent 
As is well known~\cite{fujikawa1,fujikawa2}, the fermionic measure is not invariant under local chiral transformations,
\begin{equation}
\psi(x) \to e^{i \alpha(x) \gamma_5} \psi(x), \qquad \bar{\psi}(x) \to \bar{\psi}(x)\, e^{i \alpha(x) \gamma_5}, \label{chiral}
\end{equation}
and consequently, the fermionic functional measure transforms as
\begin{equation}
\mathcal{D}\bar{\psi} \, \mathcal{D}\psi \ \to \ \mathcal{D}\bar{\psi} \, \mathcal{D}\psi \, \exp\left( -\frac{e^2}{16\pi^2} \int d^4x \, \alpha(x)\, F_{\mu\nu} \tilde{F}^{\mu\nu} \right). \nonumber
\end{equation}
\vskip 0.25cm
\noindent 
The anomalous transformation of the fermionic measure implies that local chiral rotations lead to a nontrivial contribution to the effective action, proportional to the topological density $F_{\mu\nu} \tilde{F}^{\mu\nu}$. 
To compensate for this anomaly and preserve gauge invariance at the quantum level, it is customary to introduce an additional term in the action,
\begin{equation}
S_\theta = \int d^4x\, \theta\, \frac{e^2}{16\pi^2} F_{\mu\nu} \tilde{F}^{\mu\nu},
\end{equation}
where \( \theta \) is initially treated as a constant vacuum parameter. Although this term does not affect the classical dynamics, it plays a crucial role in the quantum theory by encoding possible CP-violating effects.
\vskip 0.25cm
\noindent 
To explore more general configurations and to allow for a consistent treatment under local chiral transformations, we now promote $\theta$ to a spacetime-dependent background field $\theta(x)$. 
This generalization enables us to monitor local variations in effective action and paves the way for a deeper understanding of how topological structures influence low-energy physics. 
As we will show, this formulation naturally reveals additional structure that is not apparent at the classical level.
\vskip 0.25cm
\noindent 
Collecting all terms, the effective quantum action becomes \cite{gamboa1}
\begin{equation}
S_{\text{eff}} = \int d^4x \left( \frac{1}{4} F_{\mu \nu} F^{\mu \nu} 
+ \frac{e^2}{16\pi^2} \left( \theta(x) + \alpha(x) \right) F_{\mu \nu} \tilde{F}^{\mu \nu} 
+ \bar{\psi} \left( \slashed{D} + \gamma_5 \slashed{\alpha} \right) \psi \right).
\label{effe1}
\end{equation}
\vskip 0.25cm
\noindent 
{\bf Adiabatic approximation}: We use this action as the starting point to integrate the fermionic degrees of freedom and implement the adiabatic approximation at the level of the eigenvalue equation.
\begin{equation}
\left( \slashed{D} + \gamma_5 \slashed{\alpha} \right) \varphi_n = \lambda_n \varphi_n.
\end{equation}
\vskip 0.25cm
\noindent 
Using the Berry ansatz, we find that adiabatic evolution modifies the fermionic operator via the emergence of the Berry connection:
\begin{equation}
\gamma_5 \slashed{\alpha} \ \longrightarrow \ \gamma_5 \slashed{\alpha} + \slashed{\mathcal{A}}^{(n)},
\end{equation}
where $\mathcal{A}^{(n)}_\mu = i \langle \varphi_n | \partial_\mu \varphi_n \rangle$ is the Berry connection associated with the $n$-th eigenmode.
\vskip 0.25cm
\noindent 
This shows that, in the adiabatic limit, the effective action becomes
\begin{equation}
S_{\text{eff}} = \int d^4x \left( \frac{1}{4} F_{\mu \nu} F^{\mu \nu} 
+ \frac{e^2}{16\pi^2} \left( \theta(x) + \alpha(x) \right) F_{\mu \nu} \tilde{F}^{\mu \nu} 
+ \bar{\psi} ( \slashed{D} -  \slashed{\alpha}\gamma_5 + \slashed{\mathcal{A}}^{(n)} ) \psi \right).
\end{equation}
\vskip 0.25cm
\noindent 
In this framework, $\mathcal{A}_\mu^{(n)}$ transforms under a local phase redefinition of the eigenstates,
\begin{equation}
\varphi_n(x) \to e^{i\gamma^{(n)}(x)} \varphi_n(x), \label{eigen01}
\end{equation}
as a gauge field:
\begin{equation}
\mathcal{A}_\mu^{(n)} \to \mathcal{A}_\mu^{(n)} + \partial_\mu \gamma^{(n)}(x).
\end{equation}
\vskip 0.25cm
\noindent 
To eliminate the combined term $\slashed{\mathcal{A}}^{(n)} -  \slashed{\alpha}\gamma_5$ from the effective Dirac operator, we impose the condition
\begin{equation}
\mathcal{A}_\mu^{(n)} = \gamma_5 \alpha_\mu(x). \label{gau1}
\end{equation}
The gauge fixing condition is encoded within the action, but it is subtle due to global obstructions. As shown in equation (\ref{gau1}), this implies  
\begin{equation}
\Delta \alpha = \oint_{\mathcal{C}} dx^\mu\, \gamma_5\, \mathcal{A}^{(n)}_\mu, \label{mono}
\end{equation}
where $\Delta \alpha$ denotes the accumulated chiral Berry phase. In the adiabatic formulation of QED, one naturally encounters an effective chiral connection $\mathcal{A}_\mu$, which arises from integrating fast degrees of freedom and captures the geometric phase structure of the remaining slow modes. This connection reflects how the fermionic ground state changes under adiabatic evolution and plays a role analogous to a gauge field in parameter space. 
\vskip 0.25cm
\noindent 
Formula (\ref{mono}) encapsulates subtle physical and topological features. If the effective connection $\mathcal{A}^{(n)}$ is exact—hence also closed—then the Berry phase accumulated along any closed path vanishes, and $\alpha(x)$ can be defined as a globally well-defined function. In this case, the local chiral transformation $\psi \to e^{i\gamma_5 \alpha(x)} \psi$ is well defined, and the shift $\theta(x) \to \theta(x) + \alpha(x)$ is justified. It is then natural to interpret $\theta = \bar{\theta}$ as an effective external field, and the Lagrangian
\[
\mathcal{L} = \frac{1}{4} F_{\mu\nu} F^{\mu\nu} + \bar{\theta}(x) F_{\mu\nu} \tilde{F}^{\mu\nu}
\]
describes photons interacting with an external pseudoscalar source, analogous to a classical axion background. 
\noindent
In contrast, if $\mathcal{A}^{(n)}$ is closed but not exact, then $\alpha(x)$ does not exist as a globally defined function, and the Berry phase accumulated along a closed path becomes nonzero. In this case, the phase defines a chiral monodromy, which cannot be removed by any local chiral transformation and thus represents a global, gauge-invariant, and topological contribution to the effective theory. This phenomenon, inspired by the structure of the Schwinger model, reveals a non-perturbative aspect of QED that, to the best of our knowledge, has not been previously identified. In particular, it suggests that a geometric phase—arising from adiabatic evolution in the fermionic sector—may play a physically observable role in the infrared dynamics of QED. Such a contribution cannot be captured by local shifts of the vacuum angle $\theta$, and instead reflects a genuinely non-perturbative structure in the effective action.

\vskip 0.25cm 
\noindent 
This reflects the underlying topological structure of the theory and is intimately connected to the presence of chiral anomalies.
\vskip 0.25cm 
\noindent{\bf{Simple example}}: 
In the adiabatic framework, the Berry connection associated with a given energy level can be expressed as a $U(1)$ one-form $\mathcal{A}^{(n)}_\mu(x)$. In the Schwinger model, this connection takes the form
\[
\mathcal{A}^{(n)}_\mu = \epsilon_{\mu\nu} \partial^\nu \alpha(x),
\]
where $\alpha(x)$ is a scalar field encoding the chiral rotation of the fermionic eigenstates, and $\epsilon_{\mu\nu}$ is the antisymmetric Levi-Civita symbol in two dimensions. This connection is closed by construction, but not necessarily exact. The Berry phase accumulated along a closed path $C$ in configuration space is then given by the holonomy
\[
\oint_C dx^\mu\, \mathcal{A}^{(n)}_\mu = - \oint_C dx^\mu\, \epsilon_{\nu\mu} \partial^\nu \alpha(x),
\]
which, via Stokes' theorem, becomes
\[
\oint_C \mathcal{A}^{(n)} = \int_\Sigma \Box \alpha(x) \, d^2x,
\]
where $\Sigma$ is a surface bounded by $C$, and $\Box$ is the Euclidean Laplacian. This expression shows that the Berry phase is sensitive to the topological content of $\alpha(x)$ and reflects a global chiral monodromy. It also connects naturally to the index of the Dirac operator in the presence of gauge backgrounds and axial rotations. In particular, if $\Box \alpha$ has compact support (e.g., localized sources), the holonomy becomes quantized and encodes a nontrivial topological obstruction to remove the chiral connection locally.

\vskip 0.25cm
\noindent
{\bf Some Remarks}: In four-dimensional QED, the Berry phase accumulated by the fermionic vacuum during an adiabatic evolution can be expressed as a holonomy of an effective connection $\mathcal{A}_\mu$ defined over the space of gauge backgrounds. If $\mathcal{C}$ is a closed path in the configuration space of gauge fields—parametrizing, for instance, a cyclic adiabatic variation of the external background $A_\mu(x)$—then the geometric phase is given by
\[
\Delta \alpha = \oint_{\mathcal{C}} dx^\mu\, \mathcal{A}_\mu.
\]
This phase is global, gauge-invariant, and generally nontrivial. Unlike in two dimensions, where the connection can be locally written as $\mathcal{A}_\mu = \epsilon_{\mu\nu} \partial^\nu \alpha$, in four dimensions, no such local expression exists. Nevertheless, the holonomy $\Delta \alpha$ retains a topological character and cannot be removed by any local gauge or chiral transformation. It thus acts as an emergent angle in the theory, analogous to the vacuum angle $\theta$ in QCD. As a result, the full partition function may be written as a sum over topological sectors weighted by this geometric phase,
\[
\mathcal{Z} = \sum_n e^{i n \Delta \alpha} \, \mathcal{Z}_n,
\]
where $n$ labels the winding number associated with the adiabatic cycle $\mathcal{C}$, and $\mathcal{Z}_n$ denotes the contribution from each sector. This structure encodes a genuinely non-perturbative contribution to the effective action that arises from the Berry phase associated with the fermionic vacuum in QED.

\vskip 0.25cm 
\noindent 
I am pleased to acknowledge Pierre Sikivie's valuable discussions and comments. The author also acknowledges support from DICYT (USACH), grant number 042531GR$_\_$REG.

\end{document}